\documentclass[twocolumn,prl,showpacs,showkeywords]{revtex4}
\usepackage{graphicx}
\begin{document}

\title{Electron correlations, spontaneous magnetization and momentum
density in quantum dots}

\author{A. Bansil, D. Nissenbaum, B. Barbiellini}

\affiliation{Physics Department, Northeastern University, Boston MA 02115} 

\author{R. Saniz \footnote{Present address:
Department of Physics and Astronomy, Northwestern University,
Evanston, Illinois 60208}}
\affiliation{Departamento de Ciencias Exactas, Universidad Cat\'olica
Boliviana, Casilla \#5381, Cochabamba, Bolivia}




\begin{abstract}

The magnetization of quantum dots is discussed in terms of a relatively
simple but exactly solvable model Hamiltonian. The model predicts
oscillations in spin polarization as a function of dot radius for a fixed
electron density. These oscillations in magnetization are shown to yield
distinct signature in the momentum density of the electron gas, suggesting
the usefulness of momentum resolved spectroscopies for investigating the
magnetization of dot systems. We also present variational quantum Monte
Carlo calculations on a square dot containing 12 electrons in order to
gain insight into correlation effects on the interactions between like and
unlike spins in a quantum dot.

\end{abstract}

\maketitle

\section{I. Introduction}

As the need for nano-structures for technological applications grows, the
ability to probe and understand the electronic properties of these systems
becomes of paramount importance \cite{marcus97,reimann02,jiang03}. 
In this connection, quantum dots (QDs),
which can be viewed as artificial "atoms", offer unique opportunities as a
nanoscale laboratory for investigating the behavior of small numbers of
electrons and how the interplay between correlation and confinement
effects in such systems can give rise to novel phenomenon such as
spontaneous spin polarization of the electron gas
\cite{saniz2,yaki01,zabala98,khanna}. These and related
questions have been the subject of considerable interest in the recent
literature \cite{focus24,magn_prop}.

Here we discuss how the electronic structure of QDs can be modeled
theoretically for the purpose of gaining a handle on the essential
phenomenology of their magnetic properties. The exactly solvable model
Hamiltonian introduced in Ref. \cite{saniz2} is considered first. 
The model of Ref. \cite{saniz2}
assumes a single effective interaction parameter, $U$, which gives the
energy penalty for creating a pair of electrons with opposite spins.
Despite its simplicity, this model produces considerable richness in its
behavior and, in particular, it predicts oscillations in spin polarization
with QD radius at a fixed electron density. We delineate the signature of
spin polarization in the electron momentum density (EMD), thus setting the
stage for the application of momentum resolved spectroscopies as a window 
for investigating the magnetic properties of QDs. 

Although exactly solvable many-body models are of an intrinsic interest,
it is important to understand the nature of the parameters involved in
such models via accurate first principles computations. In this
connection, we have carried out variational quantum Monte Carlo (VQMC)
calculations in the interacting electron gas. Results for a 12 electron
system confined within a square QD are presented. The computed pair
correlation functions for like and unlike spins are used to deduce the
effective value of the parameter $U$, which enters the model Hamiltonian of
Ref. \cite{saniz2}. 

An outline of this article is as follows. The introductory remarks are
followed in Section II by an overview of the model Hamiltonian formalism
of Ref. \cite{saniz2}. 
Section III presents the VQMC approach and considers the
example of a 2D square QD. Section IV presents a few concluding remarks.

\section{II. A model Hamiltonian for Quantum Dots}

Insight into properties of QDs can 
be obtained by considering the
relatively simple model Hamiltonian \cite{saniz2}
\begin{eqnarray}
\label{eq_ham}
\hat H
&=&
\hat H_0+\hat H_1\\
\nonumber
&=&
\sum_{\nu\sigma}\epsilon^0_\nu 
a^\dagger_{\nu\sigma}a_{\nu\sigma}
+{1\over 2}U\sum_{\nu\nu'\sigma}
a^\dagger_{\nu\sigma}a^\dagger_{\nu'\,-\sigma}
a^{\phantom{dagger}}_{\nu'\,-\sigma}
a^{\phantom{\dagger}}_{\nu\sigma}, 
\end{eqnarray}
where $a^\dagger_{\nu\sigma}$ and $a_{\nu\sigma}$, respectively, are the
creation and annihilation operators for the one-particle state
$\phi_{\nu\sigma}$ with eigenvalue $\epsilon^0_\nu$.  The first term
($\hat H_0$) describes the noninteracting system. The interaction in the
second term ($\hat H_1$) 
is restricted to electrons of opposite spins.  The parameter
$U$ here can be viewed as an average energy penalty for two electrons to
possess opposite spins in the QD. It is of course energetically advantageous
for electrons to possess the same spin because that allows the
Coulomb energy to be lowered as the electrons are kept apart by the Pauli
exclusion principle.

The model Hamiltonian of form (\ref{eq_ham}) 
can be solved exactly. The
solution for the many-body wavefunction 
has the form of an
unrestricted Hartree-Fock wavefunction
\begin{equation}
\Psi= D^{\uparrow} D^{\downarrow},
\label{eq_uhf}
\end{equation}
where up and down arrows denote spin indices and
$D^{\sigma}=|\phi_{i,\sigma}({\bf r}_j)|$ is the Slater determinant formed
by one particle states $\phi_{i,\sigma}$.  Note that this solution is a
spin eigenfunction and does not suffer from the so-called spin
contamination problem \cite{saniz2}.  The total energy is given by
\begin{equation}
E=\sum_{\nu\sigma}(\epsilon^0_\nu+
{1\over 2}UN_{\sigma})f_{\nu\sigma},
\label{eq_etot}
\end{equation}
where $f_{\nu\sigma}$ denotes the Fermi occupation function.  For a given
total number of particles $N$, minimization of Eq. \ref{eq_etot} yields a
set of nonlinear equations for the populations, $N_\uparrow$ and
$N_\downarrow$, of the up and down spins, respectively.  The resulting
splitting in energy for states of opposite spin is
\begin{equation}
\Delta=U(N_\uparrow-N_\downarrow),
\end{equation}
and it is {\em uniform}, i.e., it does not depend on the quantum number
$\nu$.  The average polarization per electron is
\begin{equation}
\zeta=(N_\uparrow-N_\downarrow)/N.
\end{equation}
The degeneracy between up and down spin electrons is thus lifted and the
shell filling depends on the value of $\Delta$,
which is determined by the interaction strenght $U$.

\begin{figure}
\begin{center}
\includegraphics[width=\hsize]{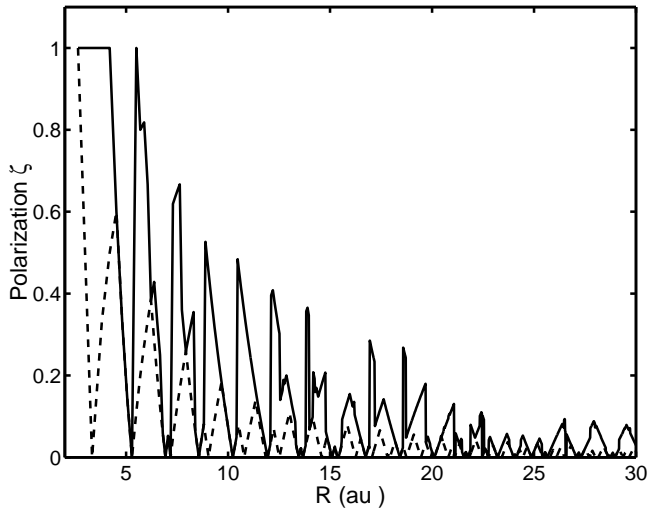}
\end{center}
\caption{Spin polarization per particle,
$\zeta=(N_\uparrow-N_\downarrow)/N$,
as a function of the dot radius $R$.
Solid line: interacting case;
dashed line: weakly
interacting case, $U\to 0$. 
Simulations are based 
on the model Hamiltonian 
of Eq. \ref{eq_ham}
using parameters
discussed in the text.} 
\label{fig_rola}
\end{figure}

Fig.~\ref{fig_rola} presents a few illustrative results
based on the model Hamiltonian
of Eq.~\ref{eq_ham}. The details of the specific 
parameters used are as follows. The
non-interacting Hamiltonian is taken to be a 3D spherical well with
potential, $V(r)=-8.62$ eV, for $r\leq R$, and $V(r)=0$, otherwise. The
electron density $n$ in the QD is kept fixed corresponding to $r_{s}$=5,
where $r_s$ is the standard parameter given by, $n (4\pi r_{s}^{3}/3)=1$.
The QD radius $R$ and the number $N$ of electrons are thus related via
$\,N=(R/r_s)^3$. $r_s=5$ gives a relatively low density, enabling
consideration of a wide range of QD radii.  The choice of $U$ is more
tricky since correlations in QDs are not well understood. However, on the
basis of arguments involving a screened Coulomb interaction, Ref.
\cite{saniz2} estimates $U= 27.13$ meV
when $N=96$ or $R=12.11$ {\AA} \cite{commentU}.

Fig.~\ref{fig_rola} shows the average
polarization $\zeta$ as a function of the QD radius $R$
for $r_s$=5. As $R$ increases and electrons are added, spin
polarization $\zeta$ reaches a peak each time a shell is half filled with
up-spin electrons and falls to zero when the shell is completed with
down-spin electrons, yielding a sequence of ``magic numbers'', i.e., $N$
values for which the QD magnetization vanishes. The oscillations in
spin polarization are damped with increasing QD size and in the high $R$
limit a paramagnetic homogeneous electron gas is recovered.
Fig.~\ref{fig_rola} also shows that the interaction
parameter $U$ can give large deviations in 
$\zeta$ from a simple
Hund's rule filling.
This is because the magnetic energy splitting 
changes with each added electron in order to
minimize the total energy given by Eq. \ref{eq_etot}.

In connection with spontaneous magnetization, it is useful to consider the
Stoner model, which is usually invoked for predicting ferromagnetism
in metals \cite{gunnar}, although it has also been applied more recently
to discuss magnetism of nanosystems \cite{zabala98}. In the Stoner
model, ferromagnetism results if
\begin{equation}
ID(\mu)\ge 1,
\label{eq_stoner}
\end{equation}
where $D(\mu)$ is the density of states (DOS) per unit cell of the
spin compensated system at the Fermi level $\mu$ and $I$ is the Stoner
parameter, which gives the gain in potential energy associated with the
occurrence of the ferromagnetic state. Within the Density Functional
Theory (DFT), $I$ can be computed using the wavefunctions of the system at
$\mu$ \cite{gunnar}. In the case of the homogeneous electron gas $I$
reduces to \cite{zabala98}
\begin{equation}
I=\frac{8[\epsilon_{xc}^F(r_s)-\epsilon_{xc}^P(r_s)]}{9(2^{4/3}-2)},
\end{equation} 
where $\epsilon_{xc}^F$ and $\epsilon_{xc}^P$ are the exchange-correlation
energy per electron in the ferromagnetic 
and the paramagnetic electron gas,
respectively. Interestingly, 
for the model Hamiltonian of Eq. \ref{eq_ham}, the
connection between the average energy penalty $U$ for having a pair of
electrons with opposite spins and the Stoner parameter $I$ is given 
as \cite{hurd81}
\begin{equation}
I=UN.
\end{equation}

Eq.~\ref{eq_stoner} makes it clear that, for a finite $I$, 
singularities
in the DOS can be expected to induce spontaneous magnetization
\cite{zabala98}.  In
QDs with high symmetry (e.g., spherical or cubic dots), symmetry
related degeneracies will generally enhance DOS peaks.
On the other hand,
symmetry breaking effects \cite{yaki01}
and disorder \cite{jacquod00}
will smear out DOS peaks and reduce
the tendency for the system to magnetize spontaneously.

\begin{figure}
\begin{center}
\includegraphics[width=\hsize]{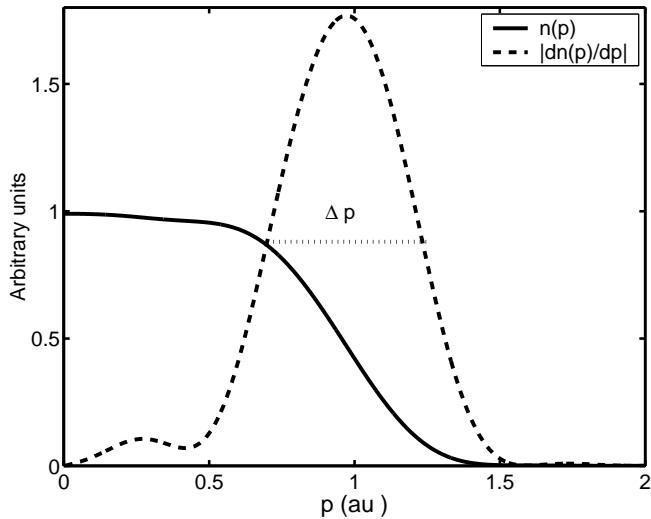}
\end{center}
\caption{
Typical EMD in a QD, $n(p)$, and the magnitude of its first
derivative, $|n'(p)|$. The position of the peak in $|n'(p)|$ defines the
QD ``Fermi momentum'', while its full-width-at-half-maximum
defines $\Delta p$.}
\label{fig_rolb}
\end{figure}

We discuss next the EMD with an eye towards identifying signatures of spin
polarization in a QD. The EMD is defined by
\begin{equation}
n({\bf p})=(2\pi)^{-1}\int_{-\infty}^\infty d\omega 
f(\omega)A({\bf p},\omega)
\end{equation} 
where $f$ is the Fermi function and
\begin{equation} 
A({\bf p},\omega)=-2{\rm Im}G^R({\bf p},\omega),
\end{equation}
is the spectral function.
The one particle Green's function $G^R(p,\omega)$ and its imaginary part
can be evaluated exactly for the model Hamiltonian of Eq. \ref{eq_ham}.
The typical behavior of the EMD and its derivative is shown in 
Fig.~\ref{fig_rolb}. The region of rapid variation in $n(p)$ can be
characterized via the position, $p_{\rm F}$, of the peak in $|n'(p)|$ and
the associated full-width-at-half-maximum, $\Delta p$.  In the bulk limit
in a metallic system, the EMD in general contains Fermi surface (FS)
breaks in the first Brillouin zone (BZ) and at the Umklapp images of the
FS in higher BZs. Correspondingly, the first derivative of the EMD
develops $\delta$-function peaks.  Although in a finite system there
cannot be breaks in the EMD, we may nevertheless refer to $p_{\rm F}$
loosely as the QD ``Fermi momentum'' for simplicity.

\begin{figure}
\begin{center}
\includegraphics[width=\hsize]{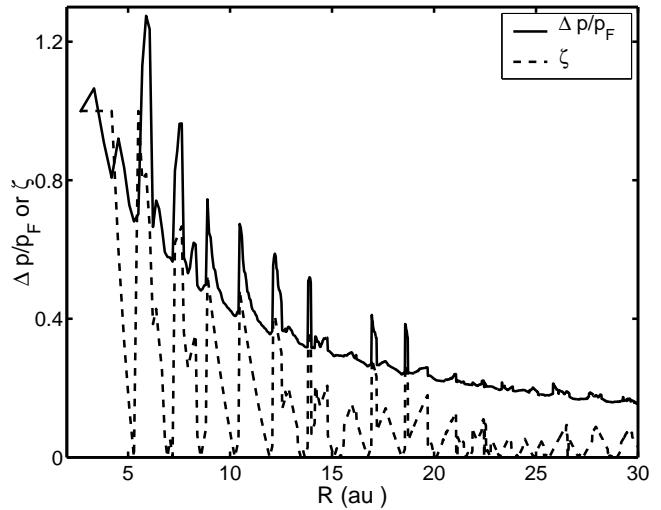}
\end{center}
\caption{
$\Delta p$ in reduced units of $p_{\rm F}$ (solid line), and
spin polarization $\zeta$ (dashed) {\it vs.} QD radius $R$.
Simulations are based on the model Hamiltonian of Eq. \ref{eq_ham}
using parameters
discussed in the text.} 
\label{fig_rolc}
\end{figure}

Fig.~\ref{fig_rolc} shows the
simulated behavior of $\Delta p$ in a QD as a function of the radius $R$
at a fixed electron density.  $\Delta p$ is seen to display peaks, which
are well correlated with those in the magnetization $\zeta$. The reason
for this correlation between $\Delta p$ and $\zeta$ is that in the
polarized system, in effect, there are two separate momentum distributions
for the up and down spin electrons. The two associated "Fermi" momenta
then give rise to two different peaks in $|n'(p)|$, which appear as
increased broadening of $\Delta p$ in the total momentum density of the
interacting system. In fact the $R$ dependence of $\Delta p$ can be
fitted as \cite{saniz2}
\begin{equation}
\Delta p/p_{\rm F}=\Delta p_0/p_{\rm F}
+c\Delta(\zeta-\zeta_0)\,R, \label{fit}
\end{equation}
where $\zeta_0$ is the spin polarization in the weakly interacting case
($U\to 0$), and $c$ is a constant, which depends on various QD 
parameters. Eq. \ref{fit} can be used to extract the polarization $\zeta$
from the measured R-dependence of $\Delta p$. These considerations
indicate that peaks in $\Delta p$ provide a distinct signature of
polarization of a QD and that spectroscopies sensitive to momentum density
can play a useful role in this connection \cite{mcp}. These results argue for
investigations of the QDs using Compton scattering and positron
annihilation experiments.

It should perhaps be noted that Compton scattering and positron
annihilation spectroscopies have developed by now into standard probes of
the EMD in materials.  Recent positron annihilation measurements on CdSe
QDs show that the effect of quantum confinement results in an increased
width $\Delta p$ of the Fermi edge in momentum space. The width $\Delta p$
seems to follow an inverse square law $1/d^2$ with particle diameter $d$
\cite{weber,icpa13}, in contrast to the $1/d$ law expected for the confined
homogeneous electron gas \cite{saniz1}.

\section{III. A Quantum Monte Carlo calculation}

The discussion of the preceding section 
is based on the Hamiltonian of Eq.
\ref{eq_ham} and it is thus limited by the form of the Hartree-Fock
many-body wavefunction given by Eq. \ref{eq_uhf}.  
We now consider a more
general many-body wavefunction and apply the VQMC approach 
\cite{harju97} to focus
particularly on understanding the nature of the effective interaction
between electrons of the same and opposite spins in a QD.  The
specific wavefunction used is  
\begin{equation} 
\Psi=J D^{\uparrow}D^{\downarrow}, 
\end{equation} 
where 
\begin{equation} 
J=\exp(\sum_{i<j}u_{ij}) 
\end{equation} 
is the Jastrow factor, which is expressed in terms of a product involving
two-body correlation factors, $u_{ij}$. The $u_{ij}$ must fulfill the
so-called cusp condition, i.e.  the singularities associated with the
kinetic energy must cancel those arising from the Coulomb potential in the
microscopic Hamiltonian. A simple form is given by \cite{harju97} 
\begin{equation}
u(r)=\frac{r}{1 + \beta r},
\end{equation}
where $\beta$ is a variational parameter.  When $J=1$ or $\beta
\rightarrow \infty$, $\Psi$ reduces to the form of an unrestricted
Hartree-Fock wave function.

The average repulsion energy between two electrons 
can be expressed as an
integral of the spin resolved radial pair correlation function
\begin{equation} 
g_{\sigma, \sigma'} \left( r \right) =
\frac{1}{nN_\sigma} \sum\limits_{i \ne j} \delta _{\sigma ,\sigma _i }
\delta _{\sigma',\sigma _j } \int \delta \left( r  - r_{ij}
\right)  \left| \Psi(\xi) \right|^2 d{\bf R} 
\end{equation}
where ${\bf R}=({\bf r}_1,...,{\bf r}_N)$,
$\xi=({\bf R},\sigma_1,...,\sigma_N)$,
$r_{ij}=|{\bf r}_i-{\bf r}_j|$,
$\delta_{\sigma,\sigma'}$ is the spin projector,
and $n$ is the average electron density.
The pair interaction energy
between two electrons of like spins 
is then given by
\begin{equation}
V_{\sigma,\sigma} = \frac{1}{{\left( {N_\sigma   - 1} \right)}}
\int {\frac{{ng_{\sigma ,\sigma } \left(  r 
\right)}}{r}d{\bf r}},
\end{equation}
and for electrons of opposite spins is
\begin{equation}
V_{\sigma,-\sigma} = 
\frac{1}{{N_\sigma  }}\int {\frac{{ng_{\sigma , - \sigma } 
\left(  r \right)}}{r}d{\bf r}}.
\end{equation}
The energy penalty
$U$ for having a pair of electrons with opposite
spins can therefore be obtained from the average
\begin{equation}
U = \frac{1}{2}\sum_{\sigma}(V_{\sigma , - \sigma }  
- V_{\sigma ,\sigma }).
\end{equation}

As an example, we have investigated a QD consisting of $12$ electrons
enclosed in a 2D square well 
of size $l=\pi$ $a_B^*$ \cite{rasanen03}. Here
and elsewhere in this section, it is convenient 
to use the modified atomic
units $a_B^*$ for length and H$^*$ for energy, which are renormalized
atomic units obtained from the effective electron band mass $m^*$ and the
dielectric constant of the material $\varepsilon$ \cite{units}.

We describe the confinement in the $xy$ plane
by an infinite hard-wall potential, therefore
the single-electron states in the square QD are
\begin{equation}
\phi_{n_x,n_y,\sigma}(x,y)=\frac{2}{\pi} 
\sin(n_x x) \sin(n_y y).
\end{equation}
For $N=12$, one obtains a closed shell system
($N_\uparrow = N_\downarrow$) with zero
net magnetization.

In the Hartree-Fock limit ($\beta \rightarrow \infty$), a Monte Carlo
calculation gives the total energy of the 12-electron QD to be $107.785
\pm 0.002$ H$^*$. The parameter $\beta$ was then optimized via the
Stochastic Gradient Approximation (SGA) \cite{harju97}. In the SGA, at
each step $n$, the value of an observable $x$ is updated with a recursive
calculation of the mean:
\begin{equation}
\bar{x}_n=\bar{x}_{n-1}-\frac{1}{n}(\bar{x}_{n-1}-x_n)~.
\end{equation} 

\begin{figure}
\begin{center}
\includegraphics[width=\hsize]{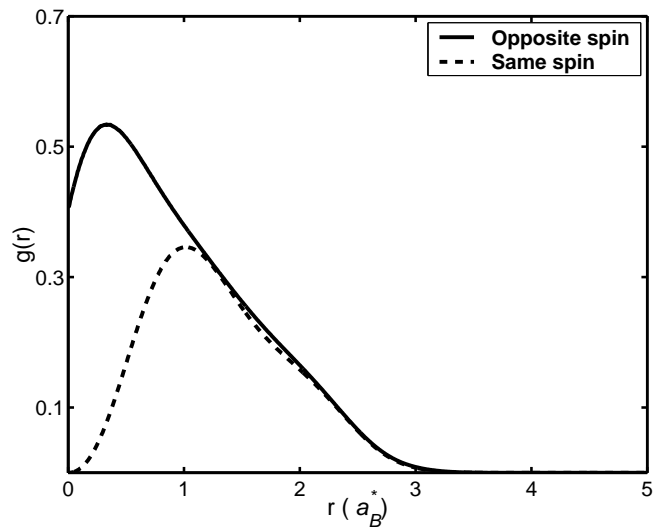}
\end{center}
\caption{Radial pair correlation functions 
$g_{\sigma, \sigma}(r)$(dashed line) and 
$g_{\sigma, -\sigma}(r)$ (solid
line) for 12 electrons 
enclosed in a 2D square well of size $l=\pi$ $a_B^*$.
}
\label{fig1}
\end{figure}

At the optimal $\beta=1.53$, the total energy is found to be $103.237 \pm
0.001$ H$^*$ with $|dE/d\beta| < 10^{-3}$ H$^*$$a_B^*$.  Using this value
of $\beta$, the results of a relatively noiseless calculation of the
spin-dependent pair correlation functions $g_{\sigma, \sigma'}$ are shown
in Fig.~\ref{fig1}.  As expected, the distribution 
$g_{\sigma, \sigma}(r)$(dashed line) 
is seen to vanish at $r=0$, reflecting the
presence of an "exchange hole" surrounding like spins due to the Pauli
exclusion principle. Electrons of unlike spins, on the other hand, still
tend to avoid each other due to Coulomb repulsion, which induces a
"Coulomb" or "correlation" hole in $g_{\sigma, -\sigma}(r)$(solid line).
Fig.~\ref{fig1} shows that the "hole," or the region of depleted electron
density, excludes a larger number of electrons and extends to a larger
distance for like spins than for unlike spins. The decrease of 
$g_{\sigma, \sigma'}$ after 
$r\approx 1$ $a_B^*$ is a geometrical effect due to the
finite size of the QD.  
The use of these correlation functions in Eqs.
16-18 yields, $V_{\uparrow \downarrow}=1.039$ H$^*$, 
$V_{\uparrow \uparrow}=0.760$ H$^*$, 
and $U=0.279$ H$^*$.  The corresponding values in
the Hartree-Fock limit are:  $V_{\uparrow \downarrow}=1.176$ H$^*$,
$V_{\uparrow \uparrow}=0.848$ H$^*$, and $U_{HF}=0.329$ H$^*$. The Jastrow
wavefunction thus leads to a reduction in the energy penalty for creating
a pair of opposite spins. The overall effect however is relatively small
in that the effective $U$ for the Jastrow wavefunction is only $15\%$
smaller than $U_{HF}$.  
These results
suggest that, despite its
simplicity, the model Hamiltonian of Eq. \ref{eq_ham}
is capable of providing a
reasonable description of the electron gas in QDs. 
Moreover, we have explicitly verified 
the oscillations of the magnetization in
the 3D-spherical quantum dots containing up to 8 electrons
by performing VQMC simulations with the SGA optimization
of the total energy.
The oscillations in magnetization with QD radius predicted on the basis of
this Hamiltonian are presumably robust to electron correlation effects
missing implicitly in the Hartree-Fock form of its solution.

\section{IV. Summary and Conclusions}

We discuss issues of spin polarization and momentum density with focus on
the treatment of correlation effects in electron gas confined within a
quantum dot. In this connection, we first consider selected results based
on a relatively simple model Hamiltonian of Ref. \cite{saniz2} 
in which interactions
are restricted via a parameter $U$ to be non-zero only for electrons of
opposite spins. This model Hamiltonian is solvable exactly and admits a
solution of the Hartree-Fock form. Moreover, it displays remarkable
oscillations in spin polarization with dot radius, which leave distinct
signature in the electron momentum density. In order to gain insight into
correlation effects more generally, we have carried out VQMC calculations
on a square dot containing 12 electrons using a Jastrow-Slater form of the
many body wavefunction. The effective $U$ value for the Jastrow-Slater
wavefunction is found to be only 15\% smaller than for the Hartree-Fock
case. On the whole, we conclude that spectroscopies sensitive to electron
momentum density--Compton scattering and positron annihilation in
particular--can potentially help delineate spin polarization effects in
quantum dots.

This work is supported by the US Department of Energy contract
DE-AC03-76SF00098, and benefited from the allocation of supercomputer time
at the NERSC and the Northeastern University's Advanced Scientific
Computation Center (ASCC).


\begin{thebibliography}{99}
\bibitem{commentU}
The Coulomb matix elements and therefore $U$ in general
depend on the size of the system.
\bibitem{units}
The band mass $m^*$ and the dielectric constant $\varepsilon$
of the polarizable background influence the modified atomic
units: H$^*$$ = (\varepsilon^2/m^*)$ H, 
$a_B^*=(m^*/\varepsilon) a_B$,
where H is the Hartree energy unit and $a_B$ is the Bohr radius.
For instance, in GaAs, using $m^*=0.067m$ and
$\varepsilon=12.4$, yields H$^*$$ = 11.8572$ meV and $a_B^*=9.79$ nm.
\end{thebibliography}
\end{document}